\documentclass[a4paper,11pt, onecolumn]{article}

\usepackage[a4paper]{geometry}		
\usepackage[font=small,labelfont=bf]{caption}

\usepackage{ae} 			
\usepackage[english]{babel} 	
\usepackage{cite}
\usepackage{amsmath} 		
\usepackage{amssymb} 		
\usepackage{bm} 			

\usepackage{float} 			
\usepackage{graphicx} 		
\usepackage{caption}
\usepackage{subcaption}
\usepackage{epstopdf}
\usepackage{sidecap}

\newcommand{\pt}{$p_{T}$}
\begin{document}

\title{Recent progress on the understanding of the medium-induced jet evolution and energy loss in pQCD}
%
%

\author{Liliana Apolin\'{a}rio 
	\thanks{\texttt{liliana.apolinario@tecnico.ulisboa.pt}} \\ 
	\textit{\small LIP Lisboa, Av. Elias Garcia 14, 1, 1000-149 Lisbon, Portugal}
}
\date{}

\maketitle
%

Motivated by the striking modifications of jets observed both at RHIC and the LHC, significant progress towards the understanding of jet dynamics within QGP has occurred over the last few years. In this talk, I review the recent theoretical developments in the study of medium-induced jet evolution and energy loss within a perturbative framework. The main mechanisms of energy loss and broadening will be firstly addressed with focus on leading particle calculations beyond the eikonal approximation. Then, I will provide an overview of the modifications of the interference pattern between the different parton emitters that build up the parton shower when propagating through an extended coloured medium. I will show that the interplay between color coherence/decoherence that arises from such effects is the main mechanism for the modification of the jet core angular structure. Finally, I discuss the possibility of a probabilistic picture of the parton shower evolution in the limit of a very dense or infinite medium.

\section{Introduction}
\label{intro}
\par Perturbative QCD (pQCD) has already proven to be very successful in describing leading particle and jet cross-sections in proton-proton (pp) collisions. However, in ultra-relativistic heavy-ion collisions, the description of such observables should account for the modifications induced by the multiple interactions with the hot and dense matter that is created in such collisions - the Quark-Gluon Plasma (QGP). The theoretical description of such modifications, experimentally observed both at RHIC and at the LHC, has been the target of extensive efforts as it provide information on macroscopic properties of the created medium and on the QCD dynamics in such dense regime. 
\par Probes originated from a hard process, with a large momentum transfer, $Q$, take place during a time and length scale $\propto Q^{-1}$ that should not be resolved by the medium. Moreover, recent results \cite{Adam2016} show that there is no strong dependency on the suppression of different hadronic species at high \pt, thus supporting a picture in which hadrons are already formed outside the medium. As such, even though the medium scale is below the $\Lambda_{QCD}$, it should be possible to describe observables related to hard processes in heavy-ion collisions within a pQCD approach. Assuming a factorisation scheme in which the medium-modifications are accounted in a modification of the parton branching structure it follows:
\begin{equation}
\begin{split}
	d \sigma_{med}^{A A \rightarrow h + \text{rest}} & = \sum_{ijk} f_{i/A}(x_1, Q^2) \otimes f_{j/A}(x_2, Q^2) \otimes \hat{\sigma}_{ij \rightarrow f+k} \otimes P_f(\Delta E, L, \hat{q}, \cdots) \\
	& \otimes D_{f \rightarrow h} (z, \mu_F^2) \, ,
\end{split}
\end{equation}
where $f_{i(j)/A}$ describe the probability of finding a parton $i(j)$ in the incoming hadron (the nuclear PDF's), $\hat{\sigma}_{ij \rightarrow j +k}$ the elementary cross-section (as in vacuum) and \mbox{$D_{f \rightarrow h} (z, \mu_F^2)$} the universal hadronisation functions from pp collisions\footnote{Even without assuming modifications to the hadronisation mechanisms in the presence of a medium, due to the presence of other colored sources, additional energy losses might be expected \cite{Beraudo2012}.} that depend on a factorisation scale, $\mu_F$ . The modifications to the showering of the parton $f$, $P_f$, will depend on the energy loss, $\Delta E$, and on the several parameters necessary to describe the medium characteristics (medium length, $L$, transport coefficient, $\hat{q}$, ...). The collection of such modifications is generically called as \textit{Jet Quenching}.
\par In this manuscript, it will be given an overview of the main results of the jet quenching description that were achieved in the last years, within a pQCD approach. In the next section, it will be given a quick overview about the main building blocks necessary to build the \textit{vacuum} parton shower (see \cite{Ellis:1991qj} and \cite{Dokshitzer:1991wu} for a complete description). In section \ref{sec:Med}, after introducing the formalism, it will be discussed the expected modifications to the elementary processes of the parton shower when in the presence of a medium, followed by a parallel discussion on the experimental evidences of the corresponding effects. Section \ref{sec:QGP} will be dedicated to an example on how to assess QGP properties using the knowledge from section \ref{sec:Med}. Finally, the conclusions will follow in section \ref{sec:Conclusions}.

\section{Vacuum Jets}
\label{sec:Vac}
\par An accurate description of the vacuum shower is of the utmost importance as it provides a well-defined baseline with respect to which we can measure modifications. Gluon bremsstrahlung is the dominant process during the shower development. The probability of emitting a gluon of energy $\omega$ and transverse momentum $k_T$ is given by:
\begin{equation}
	dP^{q \rightarrow qg} \sim \alpha_s C_R \frac{d\omega}{\omega} \frac{dk_T^2}{k_T^2} \, ,
\end{equation}
being $\alpha_s = g^2/(4 \pi)$, $g$ the coupling constant and $C_R$ the Casimir color factor ($R = F$ for parent quarks and $R = A$ for parent gluons). Two main features are evident: the presence of a lower cut-off, $k_T \simeq \theta_k \omega > Q_0 \sim \Lambda_{QCD} \Rightarrow \theta_k > Q_0/\omega$, where $\theta_k$ is the angle between the emitted gluon and the parent parton and $Q_0$ the lower virtuality scale; and the double log enhancement for collinear and soft gluon emissions. By re-summing such contributions, the total multiplicity of radiated gluons tends to be overestimated when compared to experimental observations. Thus, destructive interferences between the several emitters from a parton shower should also be taken into account. This is clearly identified when considering the emission of a soft gluon from a color neutral\footnote{The quark-antiquark pair is created by a photon.} quark-antiquark antenna of opening angle $\theta$. The spectrum of radiated gluons, from the quark, and similarly for the antiquark, after integrating over azimuthal angle, goes as:
\begin{equation}
	dN_{q}^{\omega \rightarrow 0} \sim \alpha_s C_F \frac{d\omega}{\omega} \frac{\sin \theta d\theta}{1 - \cos \theta} \Theta( \cos \theta_1 - \cos \theta) \, ,
\label{eq:vacAntenna}
\end{equation}
where $\theta_1$ is the angle between the radiated gluon and the parent parton and $C_F = (N^2 - 1)/(2N)$ being $N = 3$ the number of colours. This implies an \textit{angular ordering} between subsequent emissions such that $\theta > \theta_1 > \theta_2 > \cdots $, which effectively decreases the available phase for radiation, thus suppressing large and soft angle emissions. This can be understood setting the gluon formation time, 
\begin{equation}
	\tau_{form} \simeq \frac{\omega}{k_T^2} \, ,
\label{eq:formationTime}
\end{equation}
and its transverse wavelength, $\omega \theta_1 \simeq k_T = \lambda_T^{-1}$. During this time, $\lambda_{q \bar{q}} \approx \theta \tau_{form} = \lambda_T \theta/\theta_1$. When $\theta_1 > \theta \Rightarrow \lambda_T > \lambda_{q \bar{q}}$ and the gluon can only \textit{see} the color charge of the system\footnote{In the case of a colourful antenna - a quark-antiquark pair that is the result of a gluon splitting - there is an additional term that has in fact the opposite ordering, but whose color charge is proportional to $C_A = N$. As such, the same interpretation holds as those emissions are re-interpreted as being emitted from the initial gluon before the splitting.}. This color coherence picture was  firmly established with the results from the TASSO \cite{Braunschweig:1990yd} and OPAL \cite{Abbiendi:2002mj} collaborations. Nowadays it is the basis of a probabilistic picture of the parton shower that is implemented in every Monte Carlo event generators for collider physics.

\section{Jets in Heavy-Ion Collisions}
\label{sec:Med}
\par This section is devoted to address the main modifications to the picture depicted in the last section due to the presence of a hot and dense medium. 

\subsection{Formalism}
\label{subsec:Formalism}
\par Consider a high energy particle that propagates through an extended coloured medium. In the high-energy limit, it is assumed an ordering of scales such that $E \gg \omega \gg |k_T|, |q_T| \gg T, \Lambda_{QCD}$, where $E$ is the energy of the parent parton, $\omega$ the energy of the radiated gluon, $q_T$ and $k_T$ the respective transverse momenta and $T$ the temperature of the medium. Moreover, within a weak coupling approach, the medium is seen as a collection of independent static scattering centres such that $L \gg \lambda \gg m_{D}^{-1}$ where $L$ is the medium length, $\lambda$ the mean-free path of the particle inside the medium and $m_{D}$ the Debye screening mass that characterises the potential of the scattering centres. The medium properties are usually encoded in the so-called transport coefficient that translates the average squared transverse momentum acquired by the propagating particle per each $\lambda$, 
\begin{equation}
	\hat{q} = \frac{\left\langle k_T^2 \right\rangle}{\lambda} \, .
\label{eq:qhat}
\end{equation} 
With these assumptions, and assuming a multiple soft scattering approximation, it can be shown (see, for instance \cite{Casalderrey-Solana2007}) that such propagation, beyond the eikonal approximation\footnote{In a strict eikonal approximation, only a color phase exchange (Wilson Line, eq. \eqref{eq:WilsonLine}) is considered.}, is described by a Green's function:
\begin{equation}
\begin{split}
	G_{BA} (x_{0+}, x_{0T}, L_+, x_T | p_+) & = \int_{r_T(x_{0+})}^{r_T(L_+)} \mathcal{D} r_T (\xi) \exp\left\{\frac{i p_+}{2} \int_{x_{0+}}^{L_+} d\xi \left( \frac{dr_T}{d\xi} \right)^2 \right\} \\
	& \times W_{BA}(x_{0+}, L_+; r_T(\xi)) \, .
\end{split}
\end{equation}
By receiving transverse momentum \textit{kicks}, whose magnitude is proportional to its longitudinal momentum $p_+$, the particle undergoes Brownian motion in the transverse plane from $x_{0T}$ at the longitudinal\footnote{Light cone coordinates, $x = (x_+, x_-, x_T)$, with $x_+ = \frac{x_0 + x_3}{\sqrt{2}}$, $x_- = \frac{x_0 - x_3}{\sqrt{2}}$ and $x_T = (x_1, x_2)$, are used throughout the manuscript.} position $x_{0+}$ to $x_T$ at $L_+$. At the same time, its color field is rotated from a color state $A$ to $B$ due to the multiple interactions with the path ordered medium fields $A_-$ in the longitudinal direction, as described by a Wilson Line:
\begin{equation}
	W_{BA} (x_{0+}, L_+; x_T) = \mathcal{P} \exp \left\{ i g \int_{x_{0+}}^{L_+} dx_+ A_- (x_+, x_T) \right\} \, .
\label{eq:WilsonLine}
\end{equation}

\subsection{In-Medium Radiation}
\label{subsec:BDMPS}
\begin{figure}[h!]
\centering
\begin{subfigure}[h]{0.4\textwidth}
\includegraphics[width=\textwidth]{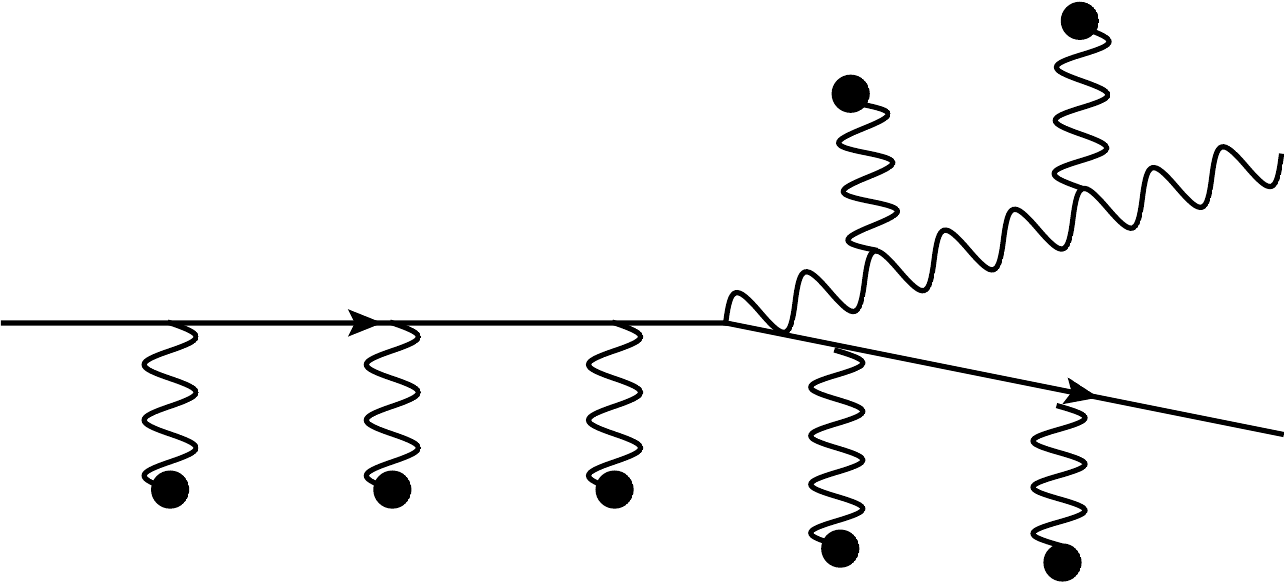}
\end{subfigure}
\hspace{0.05\textwidth}
\begin{subfigure}[h]{0.4\textwidth}
\includegraphics[width=\textwidth]{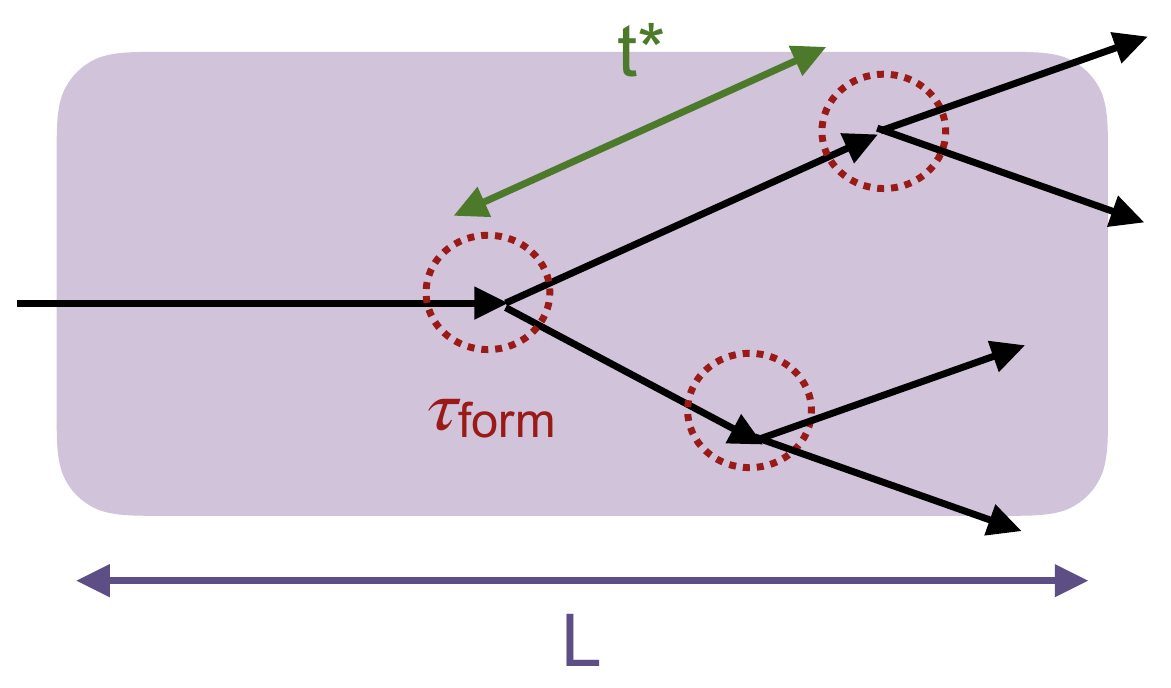}
\end{subfigure}
\caption{(Left) Schematic representation of in-medium gluon radiation that undergoes multiple soft scatterings with the medium constituents. (Right) Schematic representation of an in-medium cascade with the main timescales explicit.}
\label{fig:Schematic}
\end{figure}
\par Due to the accumulation of momenta induced by the multiple soft scattering, gluon radiation will be enhanced in the presence of a coloured medium. Such additional energy loss processes have been extensively studied in the last years (firstly in \cite{Baier1997,Baier1997a,Zakharov1996,Zakharov1997} and see \cite{Blaizot2013,Blaizot2014,Apolinario2015} for more recent works). It has been shown that the in-medium gluon radiation spectrum has two components: a factorised term that is proportional to the independent broadenings of the two outgoing particles from diagram shown in figure \ref{fig:Schematic} (left), and whose energy spectrum of subsequent independent emissions goes parametrically as
\begin{equation}
	\left. \omega \frac{dI}{d\omega}\right|_{\omega < \omega_c} \sim \alpha_s C_R \sqrt{\frac{\omega_c}{\omega}} \ \ \ , \ \ \ \left. \omega \frac{dI}{d\omega}\right|_{\omega > \omega_c} \sim \alpha_s C_R \left( \frac{\omega_c}{\omega} \right)^2 \, ,
\end{equation}
where $\omega_c =  \hat{q}L^2/2$; and a non-factorised piece in which both outgoing particles propagate coherently, suppressing the amount of energy loss with respect to the former. As shown in \cite{Blaizot2013,Apolinario2015}, the non-factorised piece is suppressed by powers of $L$. As such, the leading contribution to the gluon radiation spectrum is given by the complete factorised term. The parton shower can be considered to evolve as an incoherent sum of gluon emissions when $L \gg t^* \gg \tau_{form}$ (see figure \ref{fig:Schematic}, right), being $t^*$ the time between emissions. The timescale of an in-medium splitting, when considering equations \eqref{eq:formationTime} and \eqref{eq:qhat}, is given by $\tau_{br} (\omega) \sim \sqrt{2 \omega/\hat{q}}$. Since the number of emitted gluons is roughly given by $L/\tau_{br}$, the shower will be dominated by soft gluons. Its characteristic transverse momentum and angle are given by,
\begin{equation}
	k_{T,br}(\omega) \sim (2\omega\hat{q})^{1/4} \ \ \ , \ \ \ \theta_{br}(\omega) \sim \left( \frac{2 \hat{q}}{\omega^3}\right){1/4} \, ,
\end{equation}
thus showing that, opposite to \textit{vacuum}, in-medium gluon radiation is typically soft and emitted at large angles. Defining a critical energy for which $\tau_{br} (\omega) = L \Rightarrow \omega = \omega_c $, one finds that most of the radiation will be emitted at angles\footnote{A more careful discussion about the main differences with respect to the angular and energy characteristics of the gluon radiation spectrum between \textit{vacuum} and \textit{medium} is made in \cite{Kurkela2015,Blaizot2015}.}
\begin{equation}
	\theta > \theta_c = \frac{2}{\sqrt{\hat{q}L^3}} \, .
\label{eq:thetaC}
\end{equation}
These results show that one would expect energy loss and jet momentum broadening to appear in the jet results in ultra-relativistic heavy-ion collisions.

\subsubsection{Experimental Evidences}
\label{subsubsec:exp1}
\begin{figure}[h!]
\centering
\begin{subfigure}[h]{0.4\textwidth}
\includegraphics[width=\textwidth]{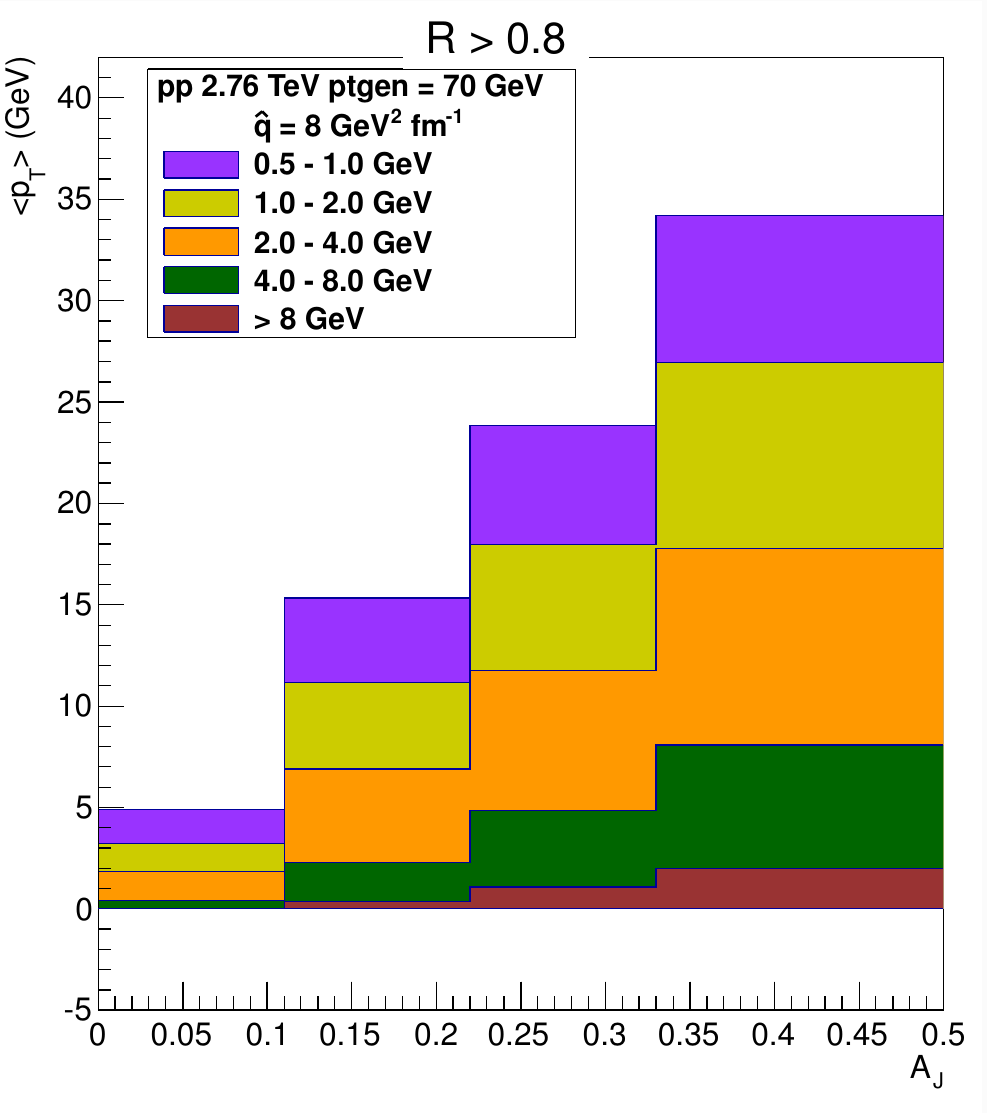}
\end{subfigure}
\hspace{0.05\textwidth}
\begin{subfigure}[h]{0.44\textwidth}
\includegraphics[width=\textwidth]{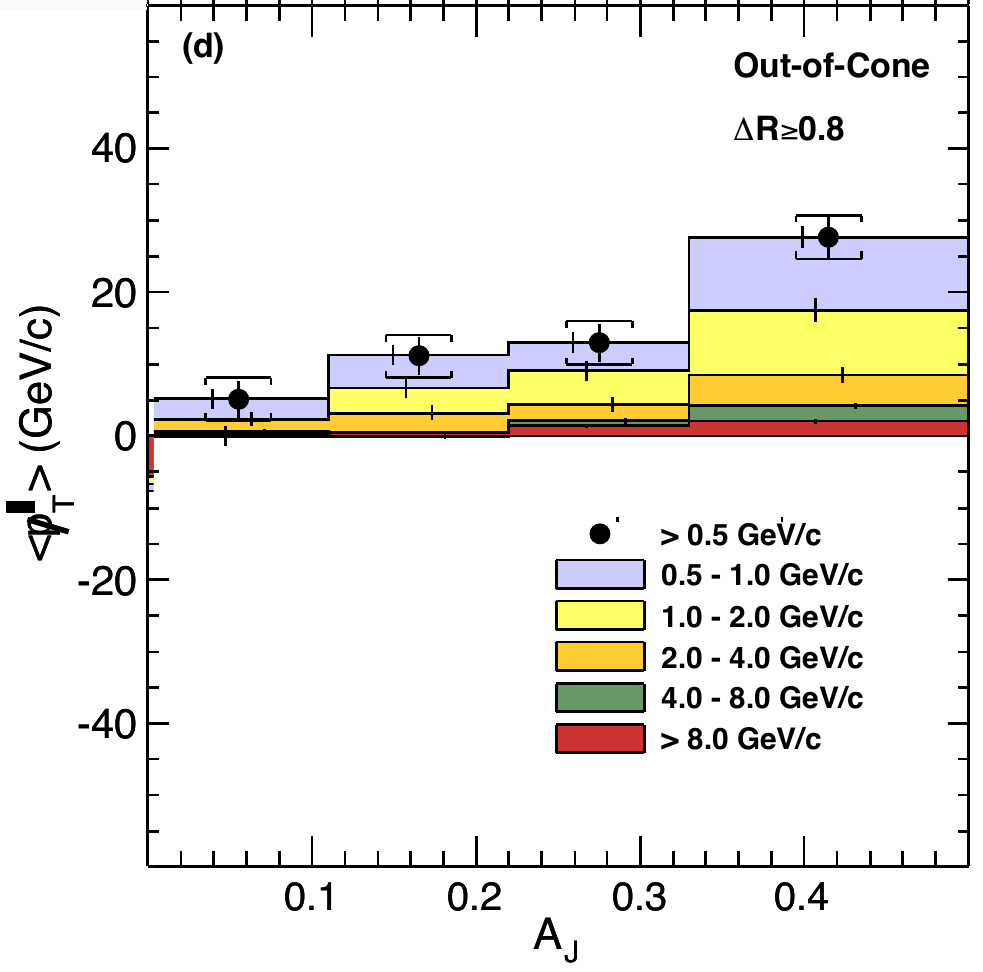}
\end{subfigure}
\caption{Results on the missing $p_T$ observable, $\left\langle p_T^\parallel \right\rangle$ at large angles with respect to the dijet axis, $R > 0.8$, for different classes of dijet asymmetry, $A_J$, for Q-PYTHIA \cite{Apolinario2013} (left) and CMS data \cite{Chatrchyan2011} (right).}
\label{fig:MissingPt}
\end{figure}
\par Q-PYTHIA \cite{Armesto2009}, a modification of the PYTHIA \cite{Sjostrand2006} event generator to account for jet quenching effects, has implemented the modifications to the gluon energy spectrum as the original works of BDMPS-Z \cite{Baier1997,Baier1997a,Zakharov1996,Zakharov1997}. A systematic comparison of this model with some experimental results was done in \cite{Apolinario2013} and it was shown to produce energy loss effects compatible to dijet measurements \cite{Chatrchyan2011}. As shown in figure \ref{fig:MissingPt}, this Monte Carlo model also produces an energy spectrum at large angles that is compatible with the CMS results on the missing $p_T$ \cite{Chatrchyan2011},
\begin{equation}
	\left\langle p_T^\parallel \right\rangle = \sum_i -p_T^i \cos ( \phi_i - \phi_{leading \ jet} ) \, .
\end{equation}
where the sum is made over all particles $i$ of the event. The dijet asymmetry, $A_J = (p_{T,1}-p_{T,2})/(p_{T,1}+p_{T,2})$, where $p_{T,1(2)}$ is the (sub)leading jet transverse momentum. These results also agree with the parametric estimates from eq. \eqref{eq:thetaC}.
\par A more recent and analytical approach to address the description of this angular broadening was carried in \cite{Blaizot2015b}. In the limit of independent emission processes, it is possible to build a parton shower based on a probabilistic picture with $t \sim L$ as the real-time evolution variable (\textit{vs} \textit{vacuum} where $t \sim \log Q^2$). In the works done by \cite{Blaizot2013a,Fister2015,Iancu2015}, this was used to explain how the energy distribution would flow as a function of time. The results show that, in the presence of a medium, the gluon emission probability is quasi-democratic. As such, the equal accumulation of gluons in all energy modes originates a constant energetic wave front that can also be used to describe the transport of energy to very large angles observed in experimental data.

\subsection{In-Medium Antenna}
\label{subsec:antenna}
\begin{figure}[h!]
\centering
\begin{subfigure}[h]{0.7\textwidth}
\includegraphics[width=\textwidth]{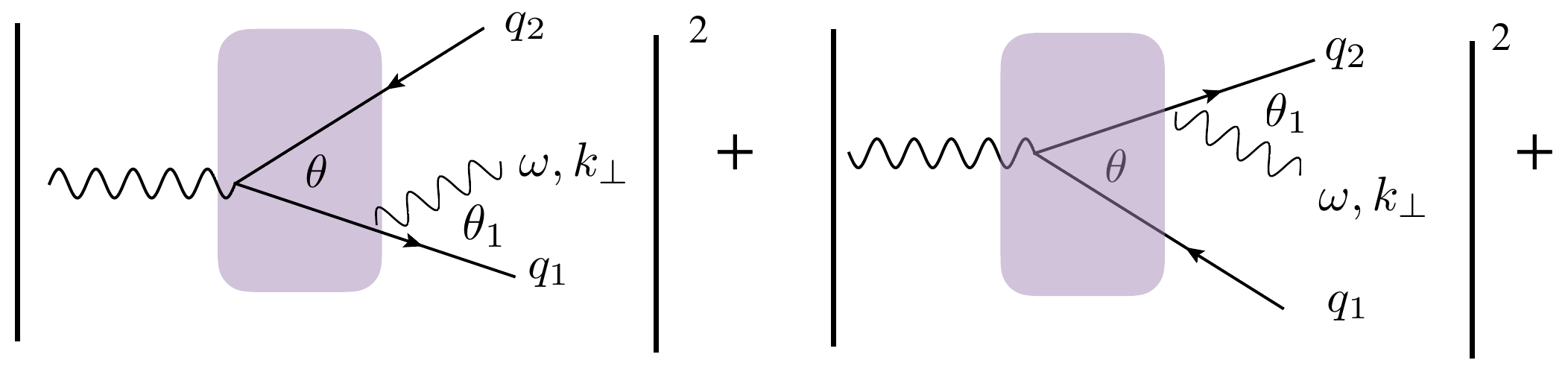}
\end{subfigure}
\begin{subfigure}[h]{0.43\textwidth}
\includegraphics[width=\textwidth]{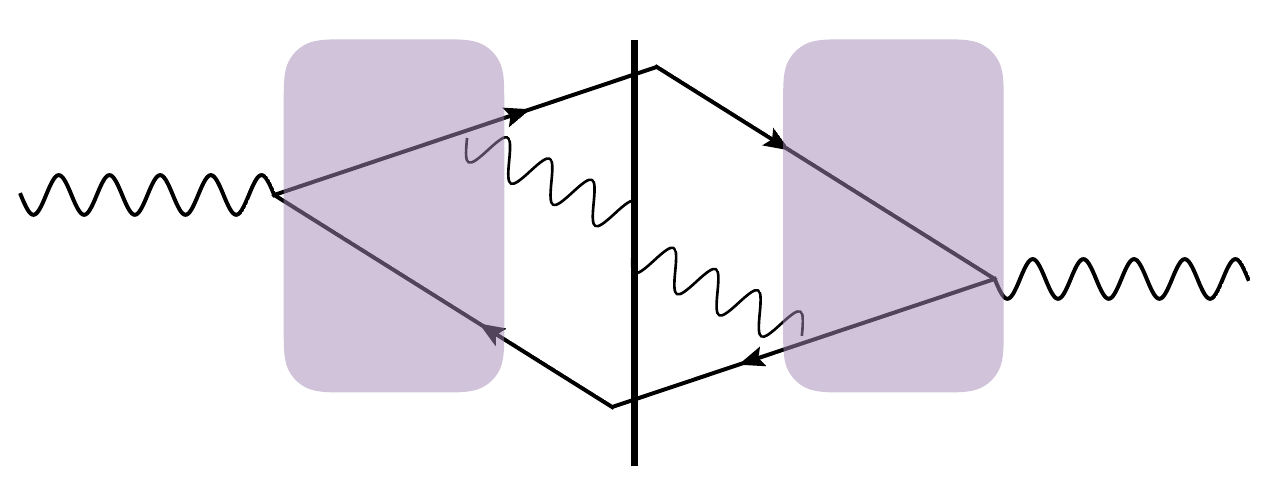}
\end{subfigure}
\caption{Diagrams that contribute to the single-gluon emission energy spectrum from a quark-antiquark antenna setup in a finite medium (represented as the purple region). The 3 contributions correspond to the sum of the first three terms in equation \eqref{eq:eikonalAntenna}.}
\label{fig:antennaDiagrams}
\end{figure}
\par To address the modifications to the radiation pattern from multiple emitters, the gluon radiation from a quark-antiquark gluon setup has been addressed, in the eikonal limit, in \cite{Casalderrey-Solana2011,Mehtar-Tani2012,Casalderrey-Solana2013}. It was shown that vacuum destructive interferences are suppressed due to color randomisation. As a consequence, the available phase space for radiation increases. This can be seen in the additional term that appears in gluon radiation spectrum, proportional to $\Delta_{med}$:
\begin{equation}
	\frac{dI}{d\Omega_k} = R_q + R_{\bar{q}} - 2 J (1-\Delta_{med}) = R_{coh} + 2 J \Delta_{med} \ \ \ , \ \ \ \Delta_{med} \approx 1 - \text{e}^{-\frac{1}{12} Q_s^2 r_T^2} \, ,
\label{eq:eikonalAntenna}
\end{equation}
where $R_q$, $R_{\bar{q}}$ and $J$ are the result of the Dirac structure from the diagrams represented in figure \ref{fig:antennaDiagrams}, $r_T = \theta L$ is the antenna transverse resolution inside the medium and $Q_s^{-1} = \left( \hat{q} L \right)^{-1/2}$ the medium transverse scale. In the soft limit, equation \eqref{eq:vacAntenna} is modified such that:
\begin{equation}
	dN_{q}^{\omega \rightarrow 0} \sim \alpha_s c_F \frac{d\omega}{\omega} \frac{\sin \theta d\theta}{1 - \cos \theta} \left[ \Theta (\cos \theta_1 - \cos \theta) + \Delta_{med} \Theta (\cos \theta - \cos \theta_1) \right] \, .
\label{eq:medAntenna}
\end{equation}
When the in-medium antenna resolution is small compared to the medium transverse scale, $r_T < Q_s^{-1}$, the medium cannot probe the quark-antiquark individually. In this limit ($\Delta_{med} \rightarrow 0$) the antenna evolves as in \textit{vacuum}, where subsequent emissions follow an angular ordering pattern. In the opposite limit, $r_T > Q_s^{-1}$, the medium can probe both coloured emitters individually and, since $\Delta_{med} \rightarrow 1$, the second term of equation \eqref{eq:medAntenna} starts to be non-negligible. The corresponding phase space for emissions opens to allow subsequent emissions to be emitted at larger angles than the previous one. This regime was denominated as \textit{anti-angular ordering} emissions as $\theta < \theta_1 < \theta_2 < \cdots$.
\par Efforts to generalize this result to account for Brownian motion in the transverse plane are currently on-going. Preliminary results show that the spectrum of radiated gluons, equation \eqref{eq:eikonalAntenna}, can be generalised such that:
\begin{equation}
	\frac{dI}{d\Omega_k} = \Delta_{coh}^\prime (R_q + R_{\bar{q}}) - 2 (1- \overline{\Delta_{med}}) J ,
\label{eq:noneikonalAntenna}
\end{equation}
where, in the collinear and soft limit, $\Delta_{coh}^\prime$ is proportional to the independent broadenings of the quark and anti-quark:
\begin{equation}
	\Delta_{coh}^\prime = \frac{4\pi}{(\hat{q} L)^2} \left[ (q_{1,T} - q_{2,T})^2 + \hat{q} L \right] \text{e}^{- \frac{ (q_{1,T}  + q_{2,T})^2 }{\hat{q} L } } \, ,
\end{equation}
and $1 - \overline{\Delta_{med}}$ is proportional to two independent harmonic oscillators:
\begin{equation}
	1 - \overline{\Delta_{med}} \sim \text{e}^{ - \frac{ i \tan (\Omega L) }{\Omega} \left[ (1-z) (q_{1,T} + k_T) - z q_{2,T} \right]^2 + \frac{ i \tan (\Omega^\prime L) }{\Omega^\prime} \left[ (1-z) q_{1,T} - z (q_{2,T} + k_T) \right]^2 } \, ,
\end{equation}
where $z$ is the fraction of energy carried by the quark with respect to the parent photon light-cone energy, $p_+$, $\Omega^2 \sim \hat{q}/[z (1-z) p_+]$ and $\Omega^{\prime 2} = -\Omega^2$. As such, the same interpretation as in the eikonal limit, equation \eqref{eq:medAntenna}, still holds.

\subsubsection{Experimental Evidences}
\label{subsubsec:exp2}
\begin{figure}[h!]
\centering
\begin{subfigure}[h]{0.6\textwidth}
\includegraphics[width=\textwidth]{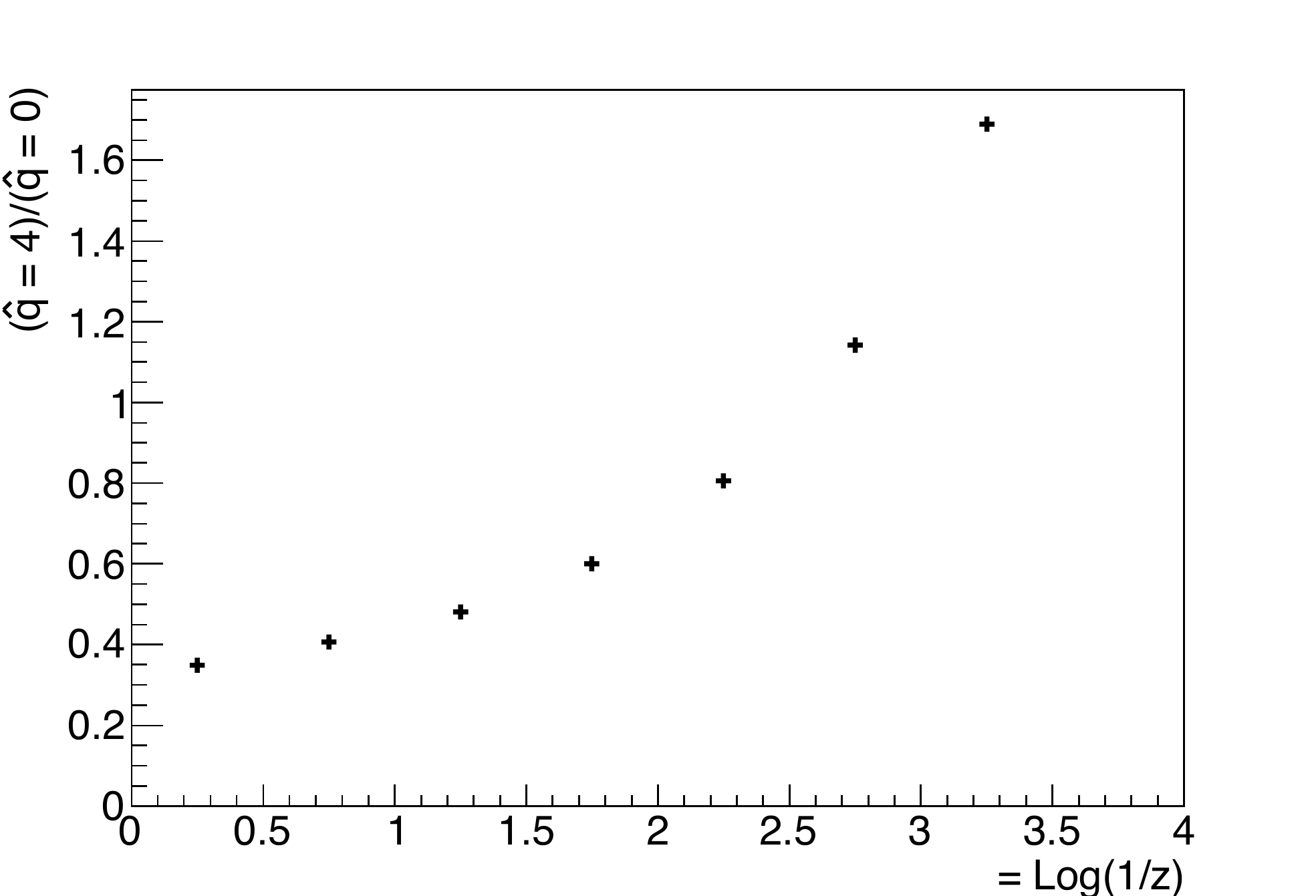}
\vspace*{1mm} 
\end{subfigure}
\begin{subfigure}[h]{0.5\textwidth}
\includegraphics[width=\textwidth]{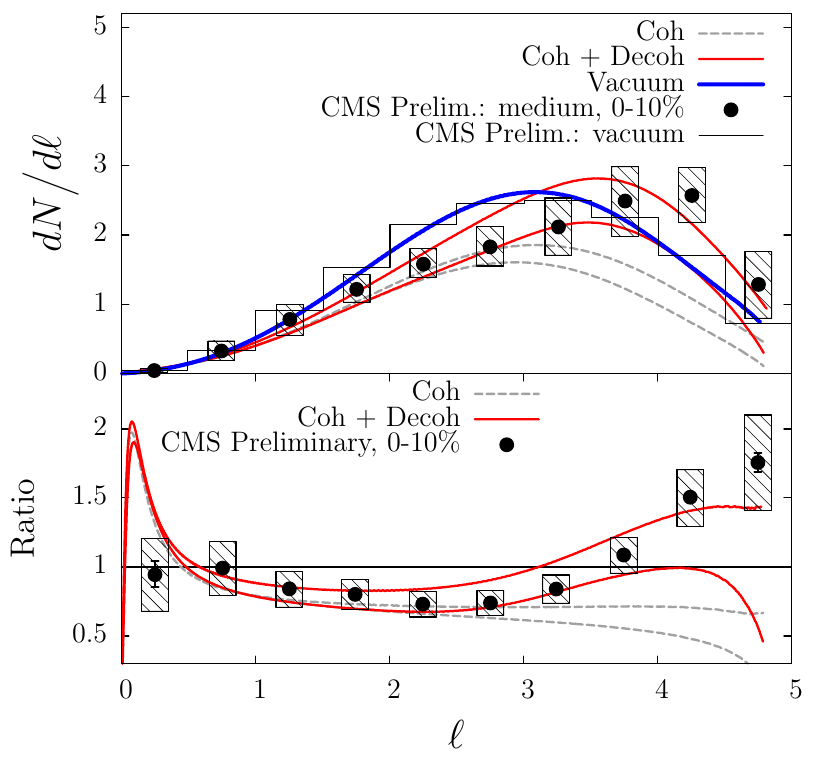}
\end{subfigure}
\caption{(Top) Results on the ratio of the fragmentation functions for Q-PYTHIA\cite{Armesto2009} using the PQM model\cite{Dainese2005} with a $\hat{q} = 4\,$GeV$^2\,$fm$^{-1}$ and $\hat{q} = 0\,$GeV$^2\,$fm$^{-1}$ (\textit{vacuum}). The data points are not made explicit but can be read from the Bottom plot. (Bottom) Results on jet multiplicity (top) as a function of $l = \log(1/x_h)$, where $x_h = \sqrt(z^2 + (m_h/p_{T,jet})^2)$, $m_h$ the hadron's mass and $p_{T,jet}$, the transverse momentum of the jet; and the ratio (bottom) of the medium-modified results (pink and dashed curves) to the reference (blue curve).}
\label{fig:jetFF}
\end{figure}
\par Results on the jet fragmentation functions\cite{ATLASCollaboration2014} and jet shapes\cite{Chatrchyan2014} show that the inner core of the jet/hard fragments are only slightly modified while the outer layers of the jet/soft fragments are enhanced with respect to pp measurements. The interplay between jet coherence/decoherence seems to be essential to get the same qualitative picture as one can observe from experimental data. Q-PYTHIA (see figure \ref{fig:jetFF}, Top) can be seen as the complete decoherence picture while in figure \ref{fig:jetFF} (Bottom) \cite{Mehtar-Tani2015}, the dashed region corresponds to the complete coherence of subsequent radiation. As one can see, none of them alone is able to describe the complete evolution of the jet fragmentation as a function of the energy carried by the particles. Nonetheless, when both phenomena are considered in the evolution of a jet, a similar trend than the one from CMS results can be achieved (pink shaded region in figure \ref{fig:jetFF}, Bottom). As such, jet coherence/decoherence phenomena might explain the observed intra-jet measurements.

\subsection{Radiative Corrections}
\label{subsec:radCorrections}
\par The probability of a high parton to acquire transverse momentum broadening is based on an instantaneous interaction with the medium (see section \ref{subsec:Formalism}). However, there are radiative corrections to the $p_T$ broadening \cite{Wu:2011kc,Liou2013} such that:
\begin{equation}
	\left\langle k_T^2 \right\rangle \simeq \hat{q}_0 L \left(1 + \frac{\alpha_s N}{2\pi} \ln^2 \frac{t}{t_0} \right) \, .
\end{equation}
The double log structure exhibits large non-local corrections that might question the validity of the probabilistic picture discussed above. These corrections can be accounted through a renormalisation of the transport coefficient yielding the same double log enhancement to the mean radiative energy loss:
\begin{equation}
	\hat{q} (t) \simeq \hat{q}_0 \left(1 + \frac{\alpha_s N}{2\pi} \ln^2 \frac{t}{t_0} \right) \ \ \ , \ \ \ \left\langle \Delta E \right\rangle \simeq \alpha_s C_R \hat{q}_0 L^2 \left(1 + \frac{\alpha_s N}{2\pi} \ln^2 \frac{t}{t_0} \right) \, .
\end{equation}
Assuming a strong ordering in the formation time of successive gluon emissions, it is possible to formulate an evolution equation \cite{Blaizot2014f,Iancu2014} that re-sums the double logarithmic power corrections:
\begin{equation}
	\frac{\partial \hat{q} (t, k_T) }{\partial \ln t} = \int_{\hat{q} t}^{k_T^2} \frac{dq_T^2}{q_T^2} \alpha(q_T) \hat{q}(t, k_T) \, .
\end{equation}
Hence, the probabilistic approximation remains valid.

\section{Assessing QGP Properties}
\label{sec:QGP}
\par Significant progress has been made to understand the physical picture of a quenched jet. The new phenomena that has been analytically established seems to in (qualitative) agreement with the experimental observations. With the current level of understanding of a medium-modified jet we are starting to be ready to follow to the next step which is to finally assess the properties of QGP.
\begin{figure}[h!]
\centering
\includegraphics[width=0.5\textwidth]{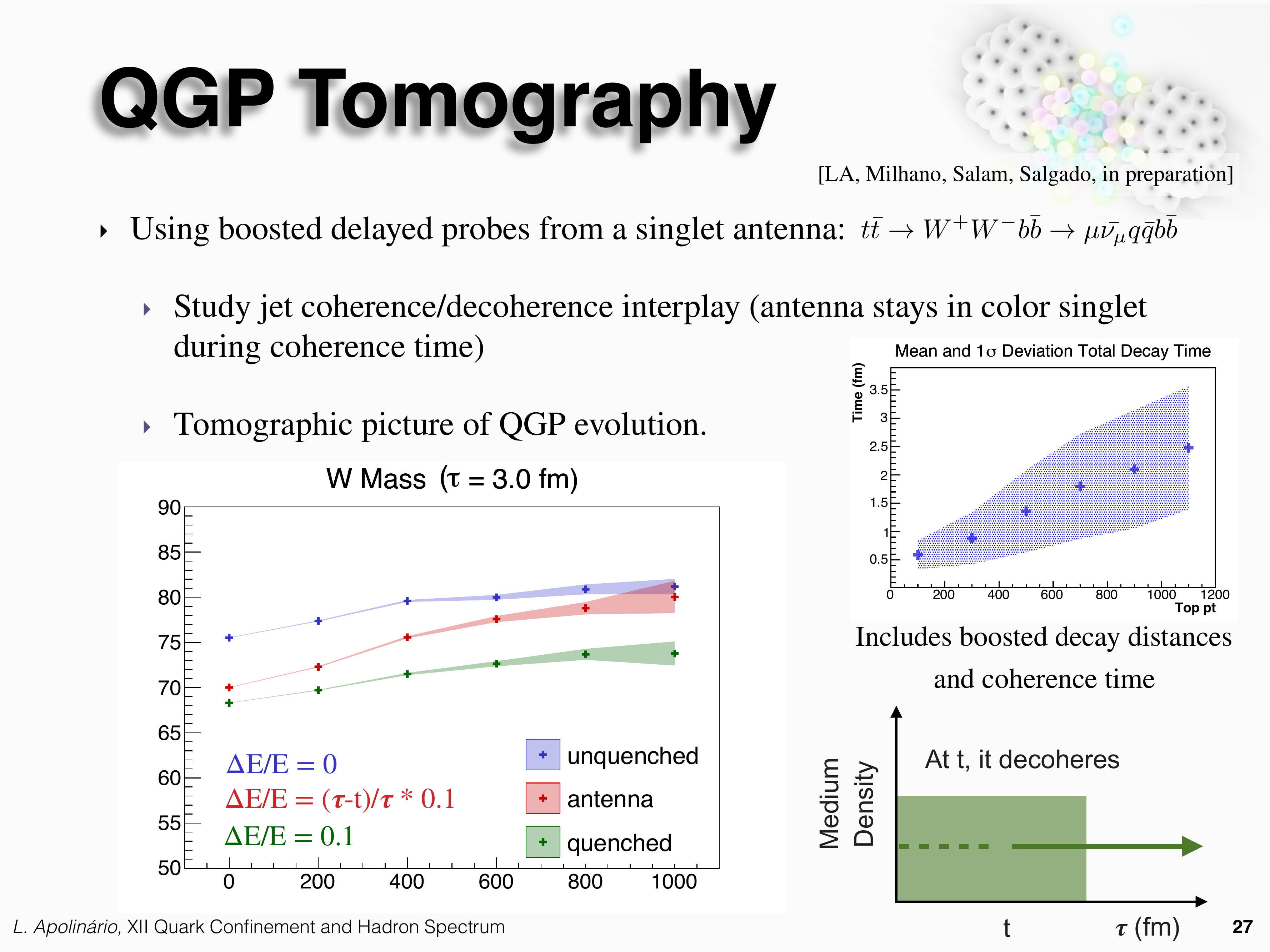}
\caption{Results on the total delay time (top decay + W decay + coherence time) as a function of the reconstructed top $p_T$. The blue dots are the average and the shaded region covers a $1\sigma$ deviation.}
\label{fig:boostedTops}
\end{figure}
\par At high energy, the production of highly boosted heavy particles is expected to be enhanced. At the time of its decay, the density profile of the QGP has already evolved. As such, by using this time delay, it should be possible to get unique insight into the time structure of the jet-QGP interaction. In the work present in \cite{Dainese:2016gch,Apolinario17}, top-antitop events were used to (1) assess the time evolution of the QGP and (2) to further explore the coherence/decoherence phenomena. The decay channel \mbox{$t \bar{t} \rightarrow W^+ W^- b \bar{b} \rightarrow \mu \bar{\nu}_\mu q \bar{q} b \bar{b}$ }was chosen as it provides a natural colourless antenna setup (the $W \rightarrow q \bar{q}$ decay) that remains in a color neutral state during a time \cite{Casalderrey-Solana2013}:
\begin{equation}
	t_{singlet} \sim \left( \frac{12}{\hat{q} \theta^2} \right)^{1/3} \, ,
\label{eq:coherenceTime}
\end{equation}
where $\theta$ is the $q \bar{q}$ opening angle. On top on this time, the top and W (boosted) decay times can be added. This will result in a window of timescales between 0.5 fm up to couple of fm that can be probed, depending on the reconstructed top $p_T$, as shown in figure \ref{fig:boostedTops}. Moreover, the $b$ jets can be used for tagging while the leptonic decay channel can provide a reference for energy loss studies\footnote{This will require a good control over the energy loss, also for $b$ quarks.}. 
\begin{figure}[h!]
\centering
\begin{subfigure}[h]{0.5\textwidth}
\includegraphics[width=\textwidth]{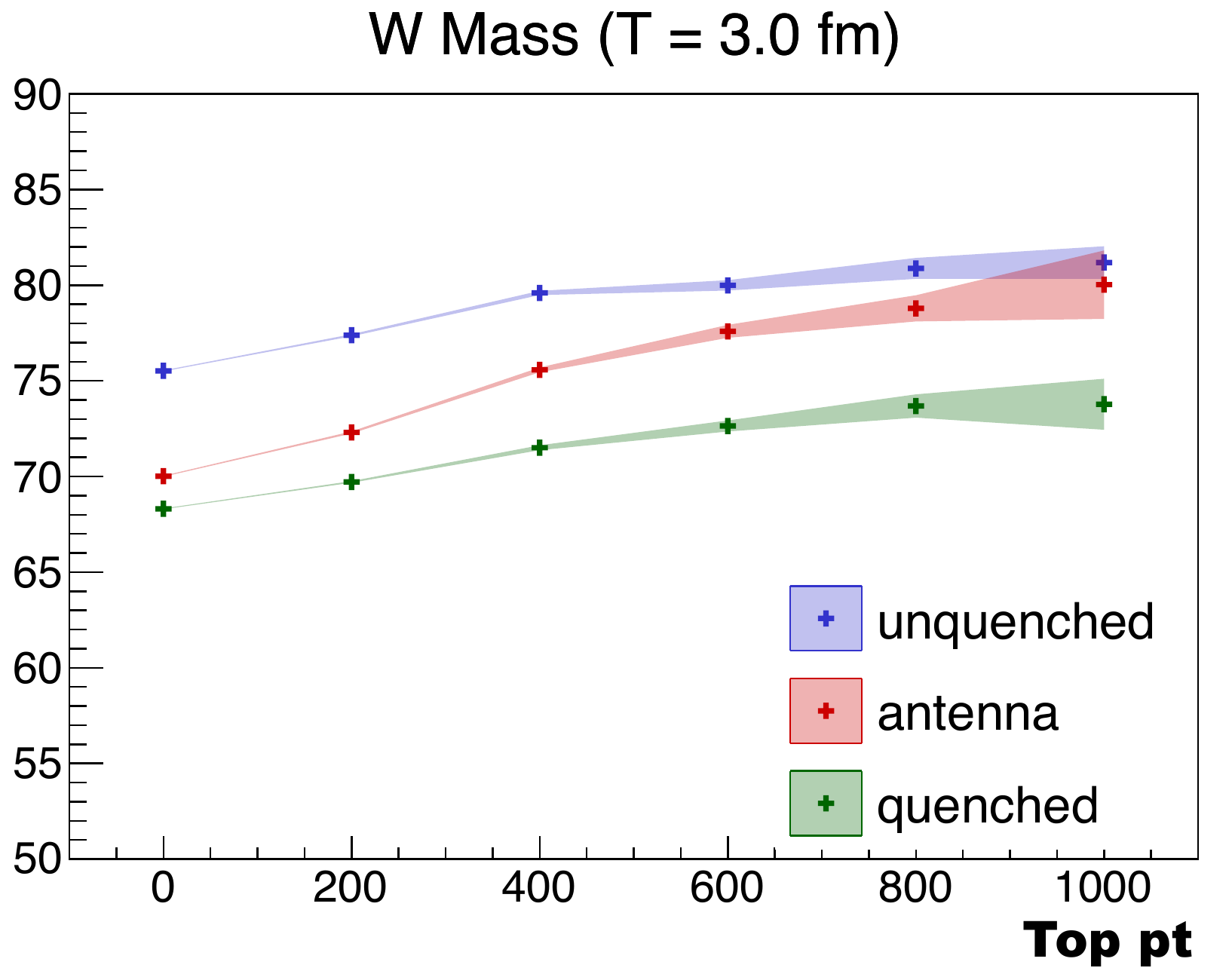}
\end{subfigure}
\hspace{0.05\textwidth}
\begin{subfigure}[h]{0.45\textwidth}
\includegraphics[width=\textwidth]{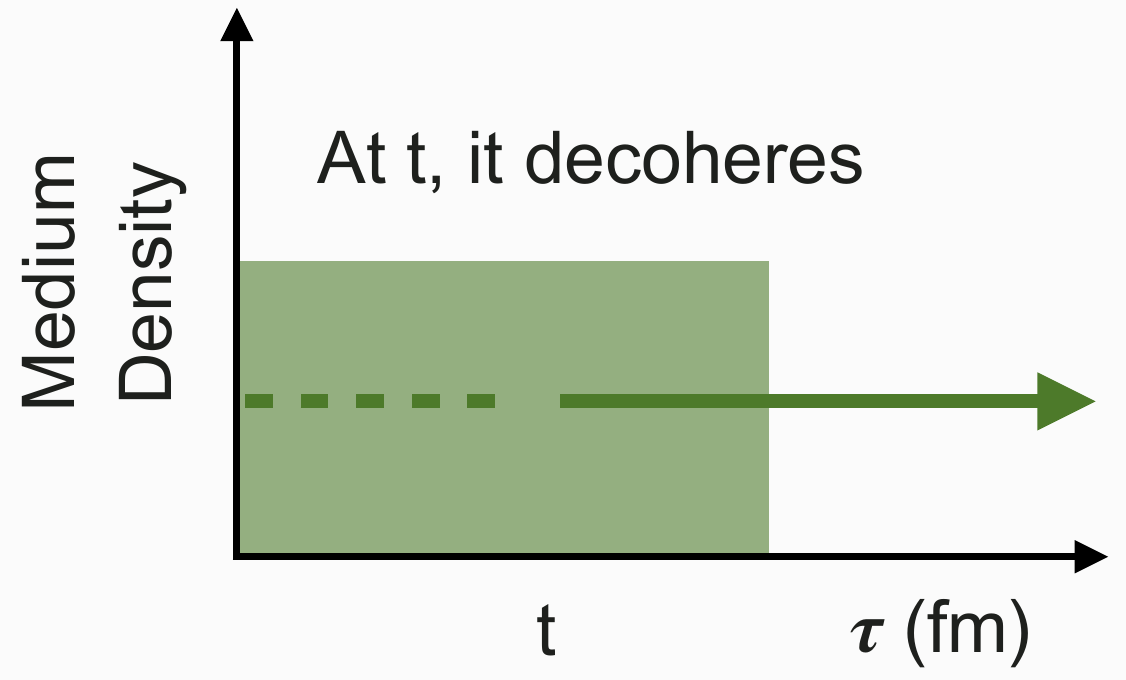}
\end{subfigure}
\caption{(Top) Reconstructed W jet mass as a function of the reconstructed top $p_T$. The dots are the average value of the obtained W jet mass and the shaded region corresponds to the estimated statistical error. (Bottom) Illustration of the model "antenna" from the left figure. The constituents from the W jet start to lose energy only after the total decay time, $t$.}
\label{fig:topsCartoon}
\end{figure}
\par By comparing the reconstructed energy to the one expected from usual energy loss processes, it should be possible to build a tomographic picture of the QGP evolution. In figure \ref{fig:topsCartoon} (Top), it is shown the reconstructed W jet mass\footnote{These results were derived for the Future Circular Collider, $\sqrt{s}_{NN} = 39\,$TeV. To calculate the available statistics, it was assumed an integrated luminosity of $30\,$nb$^{-1}$ and a $b$-tagging efficiency of $70\%$. For more details, see \cite{Dainese:2016gch}.} as a function of the reconstructed top $p_T$ for three different models: \textit{vacuum} (or pp) in blue; \textit{quenched} in green where it is assumed that all particles lose 10$\%$ of its initial energy; and \textit{antenna} where the particles will lose a fraction of energy loss that is proportional to the remaining medium length that they will travel after the total delay time (see picture from figure \ref{fig:topsCartoon}, Bottom). As one can see, very low $p_T$ jets have smaller total delay times and thus, quench more. As the $p_T$ increases, the particles will be more boosted, the decay products more collimated, and the delay time will increase. For very large $p_T$, the antenna particles will de-correlate already outside the medium experiencing no energy loss phenomena. The transition from the reference \textit{quenched} to \textit{unquenched} can give us further information on the coherence time (eq. \eqref{eq:coherenceTime}), (de)coherence phenomena and also on the fast evolution of the medium profile. Moreover, the absolute value of the reconstructed top mass is a direct probe of the integrated density profile, from $t$ to the full medium length $\tau$.

\section{Summary and Conclusions}
\label{sec:Conclusions}
\par Significant progress has been made to understand the pQCD evolution of the parton shower in the presence of a hot and dense medium. Within a pQCD picture, the jet can be understood as a collection of mini vacuum-like jets in which the angular structure follows an angular ordering pattern whenever $r_T \ll Q_s^{-1/2}$. All remaining structures, with $r_T \gg Q_s^{1/2}$, are probed independently by the medium and will lose energy, mainly in the form of soft emissions at very large angles, that will follow an anti-angular ordering.
\par One important piece that still needs to be understand is the interplay between these two showers. The ordering variable in the anti-angular ordered shower $t \sim L$ is different from the hard scale of the vacuum shower $t \sim \ln Q^2$ and there are additional scales into the problem, such as $Q_s$ and $\theta L$. Nonetheless, since the angular structure is different, it might be possible to reach a single evolution equation for both regimes to account for a full evolution of a medium-modified jet.
\par The description of the available experimental picture of a medium-modified jet is thus becoming more accurate and is starting to be used as a tool to finally probe the medium characteristics (see section \ref{sec:QGP}).
\par To further improve the current limitations of the qualitative picture obtained so far, there are on-going efforts to go beyond the current kinematical limitations of the theoretical description of the medium-modifications of the shower, in particular, the interplay between \textit{vacuum} and \textit{medium} limits discussed in section \ref{subsec:antenna}; the generalisation to multiple branchings; the study of the back-reaction;... At the same time, new observables related to the jet inner structure are being developed, such as the ones from \cite{Zhang2015,Larkoski2015}. These will help to constrain further the jet quenching mechanism and to select the final theory for medium-modifications to the QCD description of the parton shower. 
\par \
\par \textbf{Acknowledgments:} The author thanks N. Armesto and R. Concei\c{c}\~{a}o for their valuable comments and suggestions. This work was funded by the Portuguese Funda\c{c}\~{a}o para a Ci\^{e}ncia e Tecnologia, grants SFRH/BPD/103196/2014 and CERN/FIS-NUC/0049/2015.


%
\bibliographystyle{ieeetr}
 \bibliography{MyCollection.bib}

\begin{thebibliography}{10}

\bibitem{Adam2016}
{ALICE Collaboration}, ``{Centrality dependence of the nuclear modification
  factor of charged pions, kaons, and protons in Pb-Pb collisions at s NN =2.76
  TeV},'' {\em Phys. Rev. C.}, vol.~93, no.~3, p.~034913, 2016.

\bibitem{Beraudo2012}
A.~Beraudo, J.~G. Milhano, and U.~A. Wiedemann, ``{The contribution of
  medium-modified color flow to jet quenching},'' {\em J. High Energy Phys.},
  vol.~07, p.~144, apr 2012.

\bibitem{Ellis:1991qj}
R.~K. Ellis, W.~J. Stirling, and B.~R. Webber, {\em {QCD and collider
  physics}}, vol.~8.
\newblock 1996.

\bibitem{Dokshitzer:1991wu}
Y.~L. Dokshitzer, V.~A. Khoze, A.~H. Mueller, and S.~I. Troian, {\em {Basics of
  perturbative QCD}}.
\newblock 1991.

\bibitem{Braunschweig:1990yd}
W.~Braunschweig {\em et~al.}, ``{Global Jet Properties at 14-{GeV} to 44-{GeV}
  Center-of-mass Energy in $e^+ e^-$ Annihilation},'' {\em Z. Phys.}, vol.~C47,
  pp.~187--198, 1990.

\bibitem{Abbiendi:2002mj}
G.~Abbiendi {\em et~al.}, ``{Charged particle momentum spectra in e+ e-
  annihilation at $\sqrt{s}$ = 192-GeV to 209-GeV},'' {\em Eur. Phys. J.},
  vol.~C27, pp.~467--481, 2003.

\bibitem{Casalderrey-Solana2007}
J.~Casalderrey-Solana and C.~A. Salgado, ``{Introductory lectures on jet
  quenching in heavy ion collisions},'' {\em Acta Phys. Pol. B}, vol.~38,
  no.~12, pp.~3731--3794, 2007.

\bibitem{Baier1997}
R.~Baier, Y.~Dokshitzer, a.H. Mueller, S.~Peign{\'{e}}, and D.~Schiff,
  ``{Radiative energy loss and $p_T$-broadening of high energy partons in
  nuclei},'' {\em Nucl. Phys. B}, vol.~484, pp.~265--282, 1997.

\bibitem{Baier1997a}
R.~Baier, Y.~Dokshitzer, A.~Mueller, S.~Peign{\'{e}}, and D.~Schiff,
  ``Radiative energy loss of high energy quarks and gluons in a finite-volume
  quark-gluon plasma,'' {\em Nucl. Phys. B}, vol.~483, no.~1, pp.~291--320,
  1997.

\bibitem{Zakharov1996}
B.~G. Zakharov, ``{Fully quantum treatment of the Landau--Pomeranchuk--Migdal
  effect in QED and QCD},'' {\em J. Exp. Theor. Phys. Lett.}, vol.~63, no.~12,
  pp.~952--957, 1996.

\bibitem{Zakharov1997}
B.~G. Zakharov, ``{Radiative energy loss of high-energy quarks in finite-size
  nuclear matter and quark-gluon plasma},'' {\em Jetp Lett.}, vol.~65, no.~8,
  pp.~615--620, 1997.

\bibitem{Blaizot2013}
J.~P. Blaizot, F.~Dominguez, E.~Iancu, and Y.~Mehtar-Tani, ``{Medium-induced
  gluon branching},'' {\em J. High Energy Phys.}, vol.~01, no.~1, p.~143, 2013.

\bibitem{Blaizot2014}
J.~P. Blaizot, F.~Dominguez, E.~Iancu, and Y.~Mehtar-Tani, ``{Probabilistic
  picture for medium-induced jet evolution},'' {\em J. High Energy Phys.},
  vol.~06, no.~6, p.~075, 2014.

\bibitem{Apolinario2015}
L.~Apolin{\'{a}}rio, N.~Armesto, J.~G. Milhano, and C.~A. Salgado,
  ``{Medium-induced gluon radiation and colour decoherence beyond the soft
  approximation},'' {\em J. High Energy Phys.}, vol.~02, no.~2, p.~119, 2015.

\bibitem{Kurkela2015}
A.~Kurkela and U.~A. Wiedemann, ``{Picturing perturbative parton cascades in
  QCD matter},'' {\em Phys. Rev. Lett.}, vol.~114, no.~22, p.~222002, 2015.

\bibitem{Blaizot2015}
J.~P. Blaizot, L.~Fister, and Y.~Mehtar-Tani, ``{Angular distribution of
  medium-induced QCD cascades},'' {\em Nucl. Phys. A}, vol.~940, pp.~67--88,
  2015.

\bibitem{Apolinario2013}
L.~Apolin{\'{a}}rio, N.~Armesto, and L.~Cunqueiro, ``{An analysis of the
  influence of background subtraction and quenching on jet observables in
  heavy-ion collisions},'' {\em J. High Energy Phys.}, vol.~02, no.~2, p.~022,
  2013.

\bibitem{Chatrchyan2011}
{The CMS collaboration}, ``{Observation and studies of jet quenching in PbPb
  collisions at root s(NN)=2.76 TeV},'' {\em Phys. Rev. C}, vol.~84,
  no.~August, p.~024906, 2011.

\bibitem{Armesto2009}
N.~Armesto, L.~Cunqueiro, and C.~A. Salgado, ``{Q-PYTHIA: A medium-modified
  implementation of final state radiation},'' {\em Eur. Phys. J. C}, vol.~63,
  no.~4, pp.~679--690, 2009.

\bibitem{Sjostrand2006}
T.~Sj{\"{o}}strand, S.~Mrenna, and P.~Skands, ``{PYTHIA 6.4 physics and
  manual},'' {\em J. High Energy Phys.}, vol.~05, no.~05, pp.~026--026, 2006.

\bibitem{Blaizot2015b}
J.~P. Blaizot, Y.~Mehtar-Tani, and M.~A.~C. Torres, ``{Angular Structure of the
  In-Medium QCD Cascade},'' {\em Phys. Rev. Lett.}, vol.~114, no.~22, pp.~1--4,
  2015.

\bibitem{Blaizot2013a}
J.~P. Blaizot, E.~Iancu, and Y.~Mehtar-Tani, ``{Medium-induced QCD cascade:
  Democratic branching and wave turbulence},'' {\em Phys. Rev. Lett.},
  vol.~111, no.~5, pp.~3--6, 2013.

\bibitem{Fister2015}
L.~Fister and E.~Iancu, ``{Medium-induced jet evolution: wave turbulence and
  energy loss},'' {\em J. High Energy Phys.}, vol.~03, no.~3, p.~082, 2015.

\bibitem{Iancu2015}
E.~Iancu and B.~Wu, ``{Thermalization of mini-jets in a quark-gluon plasma},''
  {\em J. High Energy Phys.}, vol.~10, no.~10, p.~155, 2015.

\bibitem{Casalderrey-Solana2011}
J.~Casalderrey-Solana and E.~Iancu, ``{Interference effects in medium-induced
  gluon radiation},'' {\em J. High Energy Phys.}, vol.~08, no.~8, 2011.

\bibitem{Mehtar-Tani2012}
Y.~Mehtar-Tani, C.~A. Salgado, and K.~Tywoniuk, ``{Jets in QCD media: From
  color coherence to decoherence},'' {\em Phys. Lett. B}, vol.~707, no.~1,
  pp.~156--159, 2012.

\bibitem{Casalderrey-Solana2013}
J.~Casalderrey-Solana, Y.~Mehtar-Tani, C.~A. Salgado, and K.~Tywoniuk, ``{New
  picture of jet quenching dictated by color coherence},'' {\em Phys. Lett. B},
  vol.~725, no.~4-5, pp.~357--360, 2013.

\bibitem{Dainese2005}
A.~Dainese, C.~Loizides, and G.~Paic, ``{Leading-particle suppression in high
  energy nucleus ? nucleus collisions},'' {\em Eur. Phys. J. C}, vol.~38,
  pp.~461--474, 2005.

\bibitem{ATLASCollaboration2014}
{ATLAS Collaboration}, ``{Measurement of inclusive jet charged-particle
  fragmentation functions in Pb+Pb collisions at $\sqrt{s_{NN}}$ =2.76TeV with
  the ATLAS detector},'' {\em Phys. Lett. B}, vol.~739, pp.~320--342, 2014.

\bibitem{Chatrchyan2014}
{The CMS collaboration}, ``{Modification of jet shapes in PbPb collisions at
  sNN=2.76TeV},'' {\em Phys. Lett. B}, vol.~730, pp.~243--263, 2014.

\bibitem{Mehtar-Tani2015}
Y.~Mehtar-Tani and K.~Tywoniuk, ``{Jet (de)coherence in Pb-Pb collisions at the
  LHC},'' {\em Phys. Lett. Sect. B Nucl. Elem. Part. High-Energy Phys.},
  vol.~744, pp.~284--287, 2015.

\bibitem{Wu:2011kc}
B.~Wu, ``{On $p_T$-broadening of high energy partons associated with the LPM
  effect in a finite-volume QCD medium},'' {\em JHEP}, vol.~10, p.~029, 2011.

\bibitem{Liou2013}
T.~Liou, A.~H. Mueller, and B.~Wu, ``{Radiative pT-broadening of high-energy
  quarks and gluons in QCD matter},'' {\em Nucl. Phys. A}, vol.~916,
  pp.~102--125, 2013.

\bibitem{Blaizot2014f}
J.~P. Blaizot and Y.~Mehtar-Tani, ``{Renormalization of the jet-quenching
  parameter},'' {\em Nucl. Phys. A}, vol.~931, pp.~499--504, 2014.

\bibitem{Iancu2014}
E.~Iancu, ``{The non-linear evolution of jet quenching},'' {\em J. High Energy
  Phys.}, vol.~10, p.~95, oct 2014.

\bibitem{Dainese:2016gch}
A.~Dainese {\em et~al.}, ``{Heavy ions at the Future Circular Collider},''
  2016.

\bibitem{Apolinario17}
L.~Apolin\'ario, G.~Milhano, C.~Salgado, and G.~Salam, ``{In preparation},'' .

\bibitem{Zhang2015}
X.~Zhang, L.~Apolin{\'{a}}rio, J.~G. Milhano, and M.~P{\l}osko{\'{n}},
  ``{Sub-jet structure as a discriminating quenching probe},'' {\em Nucl. Phys.
  A}, vol.~956, pp.~597--600, 2015.

\bibitem{Larkoski2015}
A.~J. Larkoski, S.~Marzani, and J.~Thaler, ``{Sudakov safety in perturbative
  QCD},'' {\em Phys. Rev. D}, vol.~91, no.~11, 2015.

\end{thebibliography}

%
%
%

\end{document}